\documentclass[journal]{IEEEtran}
\usepackage{cite}
\usepackage{amsmath,amssymb,amsfonts}
\usepackage{algorithmicx}
\usepackage{textcomp}
\usepackage{graphicx}
\usepackage{bm}

\usepackage{amssymb}
\usepackage{mathrsfs}
\usepackage{stfloats}
\usepackage{algorithm}
\usepackage{array}
\usepackage{url}
\usepackage[colorlinks,linkcolor=black,anchorcolor=black,citecolor=black]{hyperref}
\usepackage{booktabs}
\usepackage{tabularx}
\usepackage{multirow}
\usepackage{diagbox}
\usepackage{booktabs}
\usepackage{color}
\usepackage{bbding}
\usepackage{algpseudocode}

\def\BibTeX{{\rm B\kern-.05em{\sc i\kern-.025em b}\kern-.08em
    T\kern-.1667em\lower.7ex\hbox{E}\kern-.125emX}}

\begin{document}

\title{\huge{SG-GAN: Fine Stereoscopic-Aware Generation \\
 for 3D Brain Point Cloud Up-sampling from a Single Image}}
\author{Bowen Hu, Weiheng Yao, Sibo Qiao, Hieu Pham, Shuqiang Wang, Michael Kwok-Po Ng
\thanks{Bowen Hu and Weiheng Yao contributed equally  to this work}
\thanks{Corresponding author: Shuqiang Wang, sq.wang@siat.ac.cn}
\thanks{Bowen Hu, Weiheng Yao and Shuqiang Wang are with the Shenzhen Institutes of Advanced Technology, Chinese Academy of Sciences, Shenzhen 518055, China}
\thanks{Hieu Pham is with the College of Engineering and Computer Science and the VinUni-Illinois Smart Health Center, VinUniversity, Hanoi 100000, Vietnam}
\thanks{Sibo Qiao is with the School of Software, Tiangong University, Tianjin 300387, China}
\thanks{Michael Kwok-Po Ng is with Department of Mathematics, Hong Kong Baptist University, Hong Kong}
\thanks{\textcopyright{}2025 IEEE. Personal use of this material is permitted. Permission from IEEE must be obtained for all other uses, in any current or future media, including reprinting/republishing this material for advertising or promotional purposes, creating new collective works, for resale or redistribution to servers or lists, or reuse of any copyrighted component of this work in other works.}
}

\maketitle

\begin{abstract}
In minimally-invasive brain surgeries with indirect and narrow operating environments, 3D brain reconstruction is crucial. However, as requirements of accuracy for some new minimally-invasive surgeries (such as brain-computer interface surgery) are higher and higher, the outputs of conventional 3D reconstruction, such as point cloud (PC), are facing the challenges that sample points are too sparse and the precision is insufficient. On the other hand, there is a scarcity of high-density point cloud datasets, which makes it challenging to train models for direct reconstruction of high-density brain point clouds. In this work, a novel model named stereoscopic-aware graph generative adversarial network (SG-GAN) with two stages is proposed to generate fine high-density PC conditioned on a single image. The Stage-I GAN sketches the primitive shape and basic structure of the organ based on the given image, yielding Stage-I point clouds. The Stage-II GAN takes the results from Stage-I and generates high-density point clouds with detailed features. The Stage-II GAN is capable of correcting defects and restoring the detailed features of
the region of interest (ROI) through the up-sampling process. Furthermore, a parameter-free-attention-based free-transforming module is developed to learn the efficient features of input, while upholding a promising performance. Comparing with the existing methods, the SG-GAN model shows superior performance in terms of visual quality, objective measurements, and performance in classification, as demonstrated by comprehensive results measured by several evaluation metrics including PC-to-PC error and Chamfer distance.
\end{abstract}


\IEEEpeerreviewmaketitle

\section{Introduction}

\IEEEPARstart{I}{n} recent years, with the maturity of medical technology and the environment, new surgical forms such as minimally invasive surgery have emerged in the field of brain surgery. Compared with traditional surgery, these surgical forms have characteristics of small wounds and quick recovery; but they also suffer from difficulties in visual conditions. It is often difficult for doctors to understand the pathology of the surgical area through direct observations. To bridge this gap between minimally invasive surgeries and these difficulties, a variety of technologies to assist  doctors to get supplementary visual information have been developed. Recently, some works have used intraoperative MRI (iMRI) \cite{van2018open,thomas2017novel} to improve the positioning accuracy of intraoperative navigation during minimally invasive surgery. On the basis of iMRI, \cite{hu2021point,wang2020instantiation} uses 3D reconstruction methods to generate 3D representations such as point clouds or meshes to further assist in surgical navigation. Compared with traditional 2D information, these works have proven that their results can help doctors perform surgeries better.

However, new surgical scenes put forward new requirements for these visual information assistances. The brain-computer interface (BCI) \cite{mahmood2019fully,li2018adaptive,shanechi2019brain,neural_interfaces_xu2024neural} refers to the direct connections created between the brains of humans or animals and external equipment, and completes the information exchange to achieve operating equipment. The most important factor that the brain-computer interface can accurately perform complex tasks is the installation of a large number of brain electrodes on the designated area of the operator's brain through minimally invasive surgery. \cite{liu2020neural} pointed out that with the development of brain-computer interface technology, the number of electrodes needed to detect brain signals has increased exponentially. In the current 3D reconstruction work, the point cloud is the mainstream form of the information carrier. However, current work often uses 2048 as the number of points in the point cloud. When converted to an MRI image, each point should cover about 10 $mm^3$. With the rapid increase in the number of electrodes installed in brain-computer interface surgery, a point cloud of this number of points has been unable to meet the actual needs of the operation, and it is imminent to construct a higher order of point cloud representation method.

Currently, deep learning techniques are flourishing in medical image processing. The generative adversarial networks and are also widely used in the field \cite{wang2020diabetic,yu2021tensorizing,you2020fine,hu20213d,hu2021bidirectional,8472802}, and has been extended to the 3D reconstruction \cite{30,hu2021point}. Due to the lack of high-density point cloud datasets with other existing problems, it is almost impossible to directly
generate desired target point clouds from 2D MRI images. An alternative solution is to take advantage of the more existing low-density point cloud dataset, first train a mature and high-accuracy low-density point cloud model, and then construct a network to up-sample the low-density point cloud into a high-density point cloud. The high-density point cloud should be based on the architecture and shape of the low-density point cloud to reduce the error in point cloud reconstruction.

In this paper, the stereoscopic-aware graph generative adversarial network (SG-GAN) is proposed to solve this problem; the model takes a single MRI image as input and generates a high-density, high-precision 3D point cloud. The model has two sets of generators, Stage-I GAN consists of the generator1 and discriminator1, and has a tree-structured GCN structure to generate Stage-I sparse PC by outlining the original outline and fundamental features of the object. Stage-II GAN uses a similar architecture and use stage-I results as input to reconstruct a high-density PC with detailed features.
%Because the shape of the object is already roughly described in Stage-I, Stage-II GAN only needs to correct the defects and restore the details of the target through the point cloud up-sampling process, thus avoiding the reconstruction error of direct generation.
In addition, to increase the reliability of the model reconstruction, the model should be able to extract the corresponding structural features from the image input as completely and accurately as possible. Therefore, We propose a self-attentive feature extraction encoder based on ResNet and free transforming module, which calculates the self-attentive fraction of the target without introducing additional parameters, and ensures the performance of the model while improving the computational accuracy for surgical scenarios that are sensitive to the model processing time.

\textbf {The main contributions} of this manuscript are summarized in the following.

1) A stereoscopic-aware graph generative adversarial network (SG-GAN) is proposed to reconstruct the high-density brain PC in a restricted surgical visual environment. To the best of our knowledge, this is the first work to develop an effective PC up-sampling scheme for reconstructing the 3D brain using a generative framework. A variety of new mechanisms are developed to reconstruct a promising, accurate perception of the 3D Brain shape.

2) A fine-grained perception mechanism that works in a layer-by-layer stereoscopic generating manner is designed. The mechanism enables the extraction of efficient image features and the generalization ability of two-stage generators. With the help of the multi-geometry graph convolutional neural network module, Stage-I generator can describe the branch structure of point features and outline the fundamental outline of the target.
The Stage-II generator, which contains the aggregation step, up-sampling step, and coordinate reconstruction, can enrich the object details from Stage-I and rectify possible errors.

3) A fine-grained graph neural network is designed to fit the particular scenario of 3D reconstruction. Compared to 2D images, 3D models have relationships in space that fit better with non-Euclidean data types. Our model incorporates a network construction approach based on graph convolutional networks with different geometric interests, which is better able to tap the spatiality information of the points in the point cloud.

Moreover, a free transforming module which uses a parameter-free self-attention mechanism is designed to improves accuracy while ensuring the efficiency of the model. Unlike the traditional attention mechanism that requires additional calculation parameters, it is adapted to surgical scenarios that are sensitive to model processing time. The unique spatial attention perception ability of it is also more suitable for 3D point cloud reconstruction, and has excellent feature mapping ability in large-scale image-to-point-cloud cross-domain generation.

\section{Related work}
Deep learning has become increasingly popular in the field of medical image processing\cite{10636067, 10839074}. Various kinds of deep learning technologies, including neural networks, have been applied in many medical fields, such as maturity recognition \cite{wang2018automatic,wang2018skeletal}, disease analysis \cite{wang2018bone,9130073}, cross-modal data supplement \cite{hu2020brain}, image segmentation \cite{9122459}. Many generating or reconstructing models, such as auto encoders (AEs) and generative adversarial networks, are also widely employed. There are kinds of works that incorporate these reconstructing models with 3D data\cite{hackel2016contour,wu2016learning,zhou2018voxelnet,9238491} to extend these methods from 2D image processing fields to 3D fields.

3D shape computing is a rapidly growing topic with many branches \cite{12,13,14,15,16,17}, including 3D classification and recognition \cite{18,19}, 3D reconstruction \cite{wang2018pixel2mesh,zhou2018voxelnet,20,21}, 3D completion\cite{22,25,37,28,38}, 3D upsampling\cite{li2019pu,yifan2019patch}, and so on. Recent studies have led to quick advancements in these areas. These methods aim to remove perceptual limitations, improve algorithm efficiency, provide multi-view information, complete incomplete 3D structures, or reconstruct low-density 3D inputs into high-density accurate outputs to support more complex analysis and visualization work.

\subsection{Point Cloud Generation}

The research direction of representational learning of 3D shapes focuses on reconstructing point cloud datasets as closely as possible to the real objects and guides indicative or extensional tools via reconstruction. This field has seen advancements in the form of auto-encoder models \cite{28}, variational auto-encoder models \cite{29}, and GAN architectures \cite{30,31,32} as well as diffusion models\cite{diff_luo2021diffusion} that can map a Gaussian distribution to multiple point clouds. These methods have been applied to classification experiments and fracture detection, demonstrating their validity and potential for unsupervised point cloud reconstruction.

There is a growing requirements of image-to-3D shape conversion in the medical imaging field due to the benefits from 3D shape data that enables the representation of positional and perceptual information. To address this demand, various methods are proposed to gradually improve related research. For example, \cite{33} introduced an effective loss called the geometric adversarial loss to improve the reconstruction of point cloud representations. Given a single image, PointOutNet \cite{34} can predict a 3D point cloud (PC) shape, producing point cloud coordinates directly. This deep neural network overcomes limitations of traditional methods like volumetric grids by preserving 3D shape invariance and addressing ambiguous ground-truth shapes with a novel architecture and loss function, even enabling multiple plausible predictions. Zhou et al.\cite{35} propose a method to reconstruct point clouds of organs, such as the right ventricle, from a single MRI slice using a one-stage shape instantiation method. Unlike the two-stage model in \cite{36}, which used manual segmentation and regression for liver 3D reconstruction, the approach presented by Zhou simplifies the process with PointOutNet and the Chamfer distance, achieving comparable accuracy efficiently. Further advancements include unsupervised learning, such as a study enhancing fine structure recovery (e.g., thin tubes) by matching 2D projections with irregular point supervision \cite{+2_chen2021unsupervised}, outperforming traditional methods. Similarly, Chen et al. \cite{+3_chen2024integrated} propose a cost-effective system integrating 2D image processing with neural networks for simultaneous 3D point cloud generation and segmentation, cutting training time and costs while retaining accuracy in medical and simulation contexts. These advancements demonstrate the potential of 3D shape reconstruction in the medical imaging field.

\subsection{Point Cloud Upsampling}
Currently, data-driven deep learning methods based on the learning capability of neural networks show better promise than traditional methods. Work such as \cite{18,14} makes it becomes possible to learn features directly from point clouds. In the field of point cloud upsampling, \cite{yu2018pu} proposed PU-Net, which extract multi-scale  representation of each point and extends the samples through a multi-branch fully connected layer network. \cite{yifan2019patch} proposed 3PU, which continuously repeats the input point patches in several steps and finalizes the upsampling process. 3PU is relatively computationally expensive and requires more supervision to optimize the intermediate stages of the network. Recently, \cite{li2019pu} proposed PU-GAN, a point cloud upsampling model aiming to be based on the GAN structure. PU-GAN introduces the idea of adversarial generating into this field, and the sophisticated design of multiple loss functions ensures the confidence of model convergence. \cite{qian2021pu}, on the other hand, construct a graph convolutional neural network based model to use the graph characterization of PC representation in the geometric sense for PC upsampling. To tackle sparsity, noise, and non-uniformity in point clouds, Wei et al. \cite{+4_wei2023ipunet} present a learning-based method using self-supervised cross fields aligned to sharp features, iteratively refining dense, uniform points at flexible ratios. Likewise, Zhao et al. \cite{+5_zhao2023self} introduce a self-supervised approach for upsampling, estimating projection points on implicit surfaces with two neural functions for direction and distance, enabling magnification-flexible results competitive with supervised methods.

Current methods for 3D shape reconstruction, while advanced, often struggle to capture fine spatial details and produce high-density point clouds in constrained environments like surgical settings. Our work addresses these gaps with a stereoscopic-aware graph generative adversarial network (SG-GAN), delivering precise, high-density 3D brain reconstructions through a fine-grained, layer-by-layer generation process and a graph neural network tailored to spatial relationships, offering superior accuracy and detail where existing approaches fall short.

\begin{figure*}[ht]
\centering
\includegraphics[width=18cm]{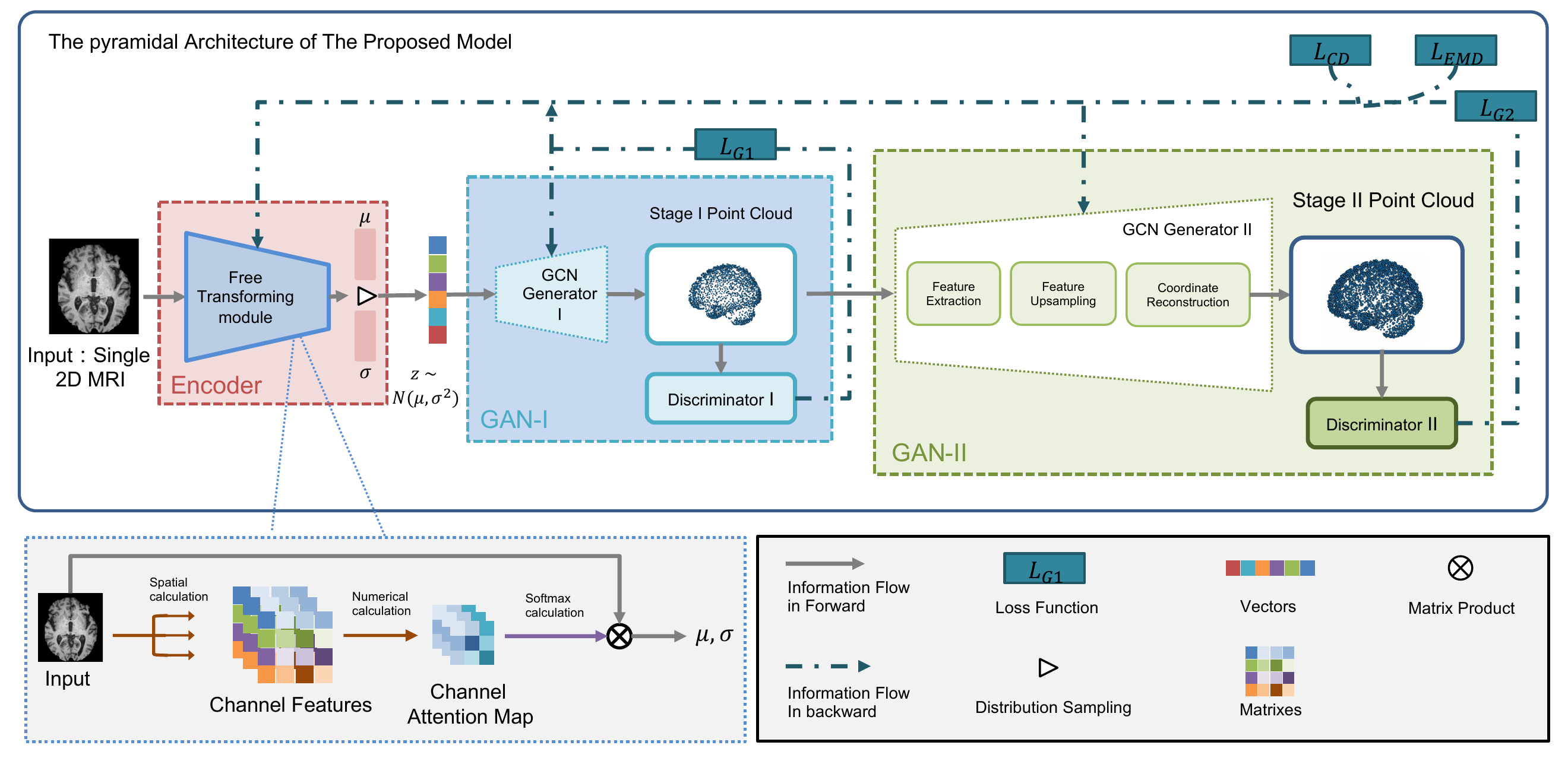}
\caption{Illustration of the SG-GAN. The encoder, composed of the free transforming module, is highlighted with pink. The stage-I GAN is marked with blue, while the stage-II GAN is marked with green. More details of several core blocks are given in the following sub-sections}
\end{figure*}

\section{Method}
The proposed SG-GAN architecture, as illustrated in Fig. 1, includes one encoder based on ResNet and the proposed free transforming module (FTM), and two GAN structures.
%Stage-I GAN generates a sparse point cloud by outlining the original shape and basic structure of the object based on the given image, while Stage-II GAN uses the Stage-I results as input to generate a high-density point cloud with details.
This approach avoids the reconstruction error of direct generation by only correcting defects and restoring details through the PC upsampling process.
A unified information flow was developed to enable communication between adjacent modules and address differences in their input and output.

\subsection{Fast feature aggregating encoder based on free transforming module}

The universal GAN framework is designed to train a generative module in which the encoder is extended to aggregate a accompanying feature representation and map the low dimension features. This mapping captures essential details about the inputs and serves as the information flow, allowing for a more comprehensive understanding of the data.

There are two key points for our SG-GAN which is aimed at the medical surgery field: a) We need to extract image features as accurately as possible and comprehensively aggregate information representing the target shape. b) Due to the intraoperative requirement for algorithmic response time, we do not want to trade off time complexity for model accuracy or introduce additional parameters. Instead, we aim to build a module based on pure numerical calculations to support tasks. Based on these requirements, we develop FTM by introducing a parameter-free self-attention mechanism into the conventional ResNet network. The proposed FTM enhances feature representation while preserving computational efficiency.

%Unlike conventional attention blocks or transformer structures, FTM aims to reduce the consumption of computational resources by avoiding the introduction of additional computational parameters in the calculation of weight scores that represent the importance of features. Instead, FTM endeavors to construct a purely numerical computational attention network and do not introduce any extra parameters to the model.

In visual neuroscience, \cite{webb2005early} noted that a different pattern of firing from the peripheral neuron or inhibition of peripheral neuron firing is usually a possible sign of high activity and information in the corresponding neuron. According to the similar guiding principles, we designed the attention-like scheme that guides deep neural networks to obtain adjusting weights. In general, our goal is to find a direct mapping method (rather than using neural networks or other large parameter structures) to build a basic attention mechanism. Therefore, FTM should only contain numerical calculations, thereby saving the necessary storage and computing costs for medical scenarios. Our model is still hierarchical, but all network architectures will be cancelled. Given a specific feature $\bm{T^k} \in \mathbb{R}^{H^k \times W^k \times C^k}$ which is located in the $k$-th layer of the ResNet, The energy function $e_t$ of a specific element $t$ in the feature tensor $T$ can be defined as
\begin{equation}
\bm{\omega} = \frac{1}{\sharp_{C^k}-1}{\sum\limits_{i=1}^{\sharp_{C^k}-1}(-1-(\omega_{t}t_i + b_t))^2},
\end{equation}
\begin{equation}
e_t = \bm{\omega} + (1 - (\omega_{t}t + b_t))^2 + \lambda\omega_{t}^2.
\end{equation}
In these equations, $\sharp_{C^k}$ is the number of elements in the channel $k$ where $t$ located in, also calculated as $H_k \times W_k$. $\omega_{t}$ and $b_t$ constitute a linear transform of $t$, it also will be applied when calculating partial coefficient of $t_i$. $\lambda$ is a regularizer.
By minimizing this energy function, one can identify the linear separability between the specific element and others in the corresponding row of the feature tensor.

Usually, we have to calculate the mean $\mu_t$ and variance $\sigma^2_t$ for each element in the channel to solve this minimizing equation, which represents all the elements in the channel except for the current element. This is computationally overly complex. Instead, we consider that all elements in the same channel obey the same or extremely similar distribution, which means that their means and variances will be close to the same. Therefore, we calculate only one mean $\mu_C$ and variance $\sigma^2_C$ for each channel. Then, in a specific channel, we can use this group of results to calculate all the elements. The mean and variance can be calculated by
\begin{equation}
\mu_C = \frac{1}{\sharp_{C^k}}\sum\limits_{i=1}^{\sharp_{C^k}}t_i
\end{equation}
\begin{equation}
\sigma^2_C = \frac{1}{\sharp_{C^k}}\sum\limits_{i=1}^{\sharp_{C^k}}(t_i - \mu_C)^2
\end{equation}

Thus, we give the attention score of a specific element $t$ as
\begin{equation}
a_t = \frac{4(\sigma^2_C + \lambda)}{(t - \mu_C)^2 + 2(\sigma^2_C + \lambda)}.
\end{equation}

\begin{figure}[ht]
	\centering
	\includegraphics[width=8.8cm]{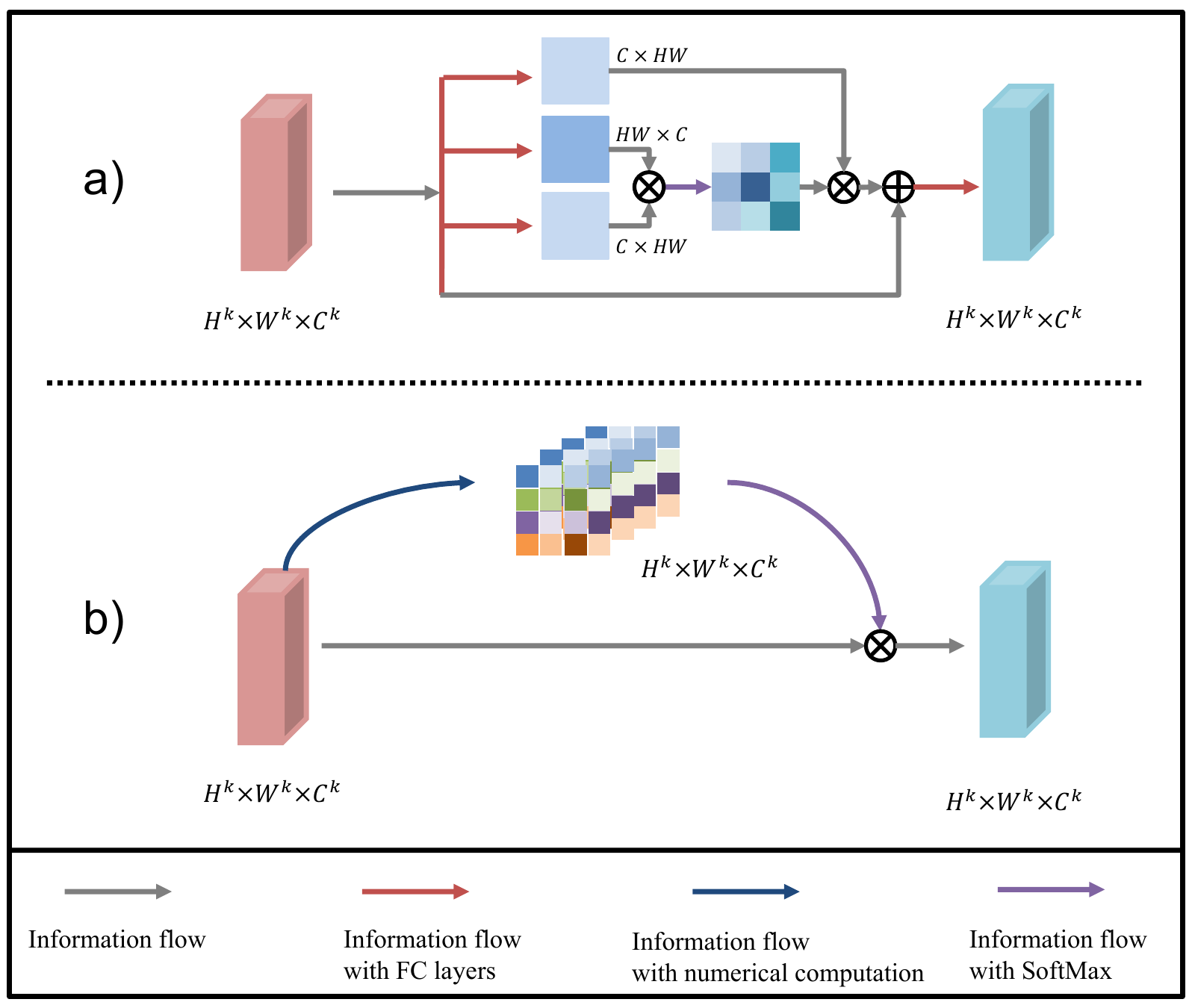}
	\caption{Comparison of a) the traditional self-attention mechanism and b) the proposed FTM.}
\end{figure}

\begin{figure*}[ht]
\centering
\includegraphics[width=0.9\textwidth]{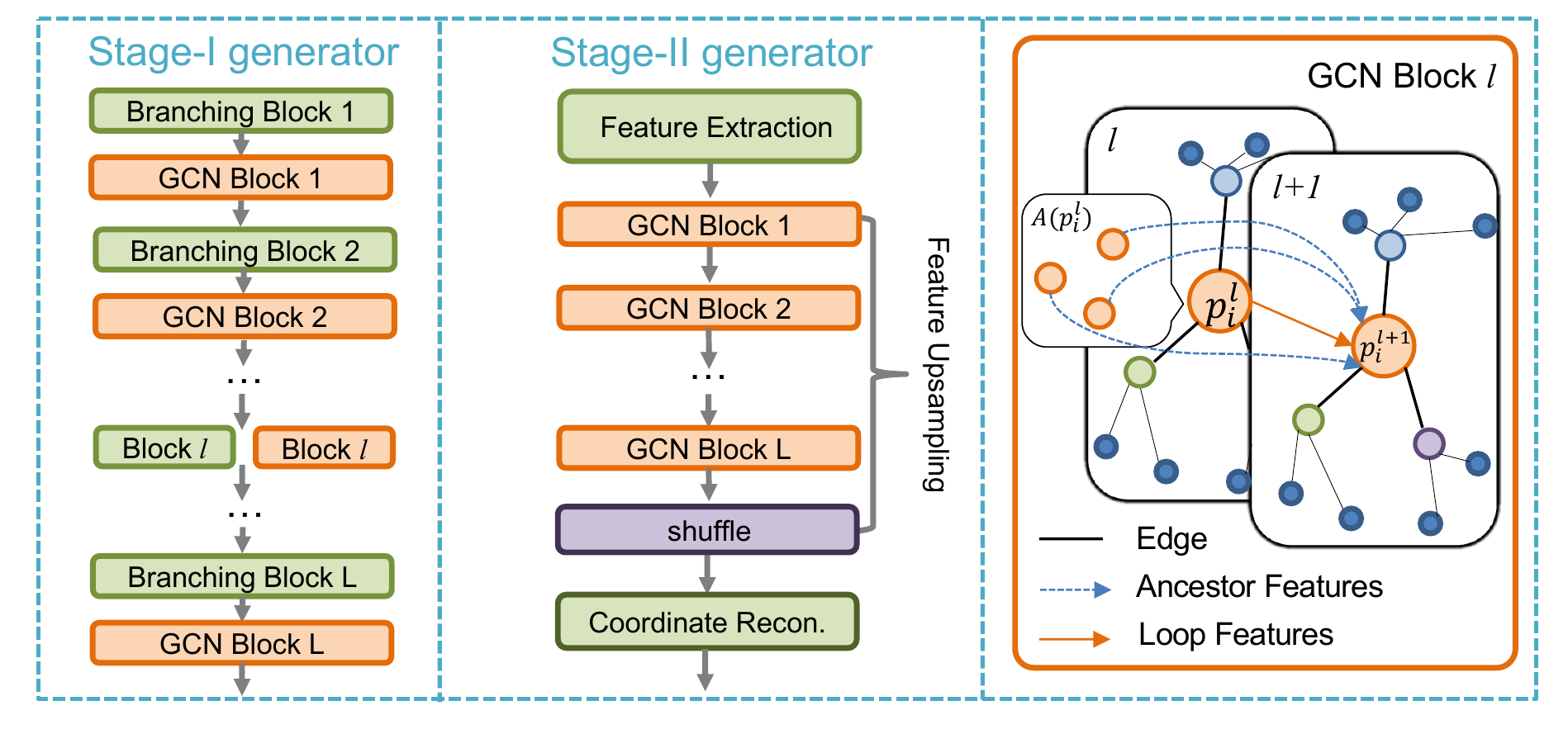}
\caption{Illustrations for the stage-I and stage-II generators. The stage-I generator based on the multi-geometry graph convolutional neural network describe the branch structure of point features and outline the fundamental outline of the target. The stage-II generator acquire the results from Stage-I and generates high-density point clouds with detailed features and it is capable of correcting defects and restoring the detailed features of the region of interest through the up-sampling process.}
\end{figure*}

%The more general meaning of the attention score is that the higher the score, the more important the element is, which does not meet the minimization goal of the current definition of attention score. Therefore, we unify the model of the attention mechanism by defining the reciprocal of the current attention score as the new attention score. It can be described as

To achieve the minimization goal, we redefine an updated attention score as:

\begin{equation}
a_t = \frac{(t - \mu_C)^2 + 2(\sigma^2_C + \lambda)}{4(\sigma^2_C + \lambda)}.
\end{equation}

%Through the calculation of the attention score for each element, we can gradually derive an attention score map $A^k$, which is akin to the conventional attention mechanism. However, our attention score map is defined for each element, rather than just channel-wise or spatial-wise, making it more responsive to subtle microstructural changes in medical images. Once we have obtained the complete attention score map, we adjust the output weights of our network layer $k$ using the formula

Further, the output weights of the network layer $k$ can be adjusted by

\begin{equation}
\bm{T^k} \gets softmax(A^k) \times \bm{T^k}.
\end{equation}
We incorporate this scheme after the middle layer to adjust the model element by element, thereby enhancing its accuracy.
%We use this network as a framework to construct an FTM and build our encoder. As a result, our encoder can extract features from the input image in a highly perceptive manner.
Unlike the conventional attention module, the proposed FTM does not employ any extra parameters to calculate attention scores, all translations are performed through numerical computation. This approach ensures temporal sensitivity in the surgical domain. A comparison between the traditional attention network and the proposed attention module is illustrated in Fig. 2.

\subsection{Stage-I GAN: branching with tree-structured GCN}

For 3D reconstruction, the first step is to restore the physical features of MRI samples, which contain rich structural information. An encoder based on FTM is used to extract geometric information and reconstruct the corresponding PC that can capture the original structure features as much as possible.
With a specific mean $\mu$ and standard deviation $\sigma$, the encoder can generate vectors $z \in \mathbb{R}^{96}$ through a Normal distribution as outputs.
Then, a proper 3D shape based on a point cloud should be partially reconstructed by the generator of SG-GAN. For a specific sample, its PC can be represented by a matrix $Y_{N\times3}$.
%where each row vector represents the 3D coordinate of a vertex.
The goal of the stage-I GAN is to generate stage-I low-density point clouds $Y^l_{2048\times3}$ and refine the overall structure of $Y^l$ as much as possible to outline the reconstructed brain. The stage-I generator takes $z$ as input and is constructed as a tree structure based on a series of basic GCN module, as well as branching module, to adjust the point clouds. Finally, the stage-I discriminator differentiates the outputs of the generator and real point clouds and optimises the generator, similar to WGAN-GP \cite{gulrajani2017improved}.

\subsubsection{The basic GCN block}
To accurately represent the intricate brain structure, the generator requires robust spatial awareness to precisely locate each point. However, the dimensionality reduction characteristic of FC-layers destroys potential contact information between adjacent elements, i.e., between adjacent points with the same dimension coordinates. Also, CNN's efficiency is limited to traditional Euclidean data. To avoid these disadvantages, GCN is used in this work to construct the generator, which is better suited for generating point clouds. Various groups consisting of branching blocks and basic GCN blocks are used to generate the 3D shape and enhance the generating details. The basic GCN block can be described as follows:
\begin{equation}
p^{l+1}_i = \sigma \left (\bm{F}^{l}_{K}(p^{l}_i) + \sum\limits _{q_j \in A(p^{l}_i)}U^l_jq_j + b^l\right ).
\end{equation}
The main focus of this formula revolves around three terms that require consideration: the loop term $S^{l+1}_i$, the ancestor term $U^{l+1}_i$, and bias $b^l$. $ \sigma (\cdot)$ describes an activation function.

The loop term, which can be expressed as
\begin{equation}
S^{l+1}_i = \bm{F}^{l}_{K}(p^{l}_i),
\end{equation}
enables the subsequent layers to possess the ability to perceive point features obtained from the upper layer. Unlike conventional graph convolutional networks that use a single parameter matrix $W$, a K-support fully connected layer $\bm{F}^{l}_{K}$ is used in the loop term to represent a more accurate distribution. $\bm{F}^{l}_{K}$ using K nodes $(p^l_{i, 1}, p^l_{i, 2}, ..., p^l_{i, k})$, to ensure a strong fitting similarity in the large graph calculation.

In conventional GCN, the neighbor term is used to propagate features from neighboring vertices to the corresponding next connected vertex. However, for the current work, PC can be reconstructed continuously via the mapping one single vector which is calculated as a high dimension point, resulting in unknown connectivity of the computational graph. The ancestor term is used instead of the neighbor term to propagate features from the ancestors of a point to the immediately following point of it. The modification allows features to be propagated more accurately and effectively in the absence of known connectivity information. This term can be described as
\begin{equation}
U^{l+1}_i = \sum\limits _{q_j \in A(p^{l}_i)}U^l_jq_j,
\end{equation}
to guarantee the inheritance of structural information and the generating ability of multiple classes of point clouds. $A(p^{l}_i)$ is the ancestors set of the point $p^{l}_i$. These ancestors employ a linear mapping matrix $U^l_j$ to map feature spaces from the other channels to $p^{l}_i$ and aggregate information to it."

\subsubsection{Tree-structure GCN based on branching blocks}
A branching process is employed to map a single point to multiple points with kinds of branch modules.
%with different branching degrees $(d_1, d_2, ..., d_n)$ utilized in various branching blocks.
Specifically, given a point $p^{l}_i$, the branching process generates $d_l$ points, resulting in a point set size that is $d_l$ times that of the upper layer.
%By controlling the degrees of branching, we ensure that the reconstruction output in this work is accurate at 2048 points.

We design branching blocks to construct the GCN to construct the stage-I generator together with the basic GCN module. The main task of the stage-I GAN is to map a feature vector into a low-density point cloud describing the subject contour, and its generation process can be roughly viewed as a tree unfolding process. Therefore, we design this stage of GCN as a tree topology to complete this process as well.

Continuous transfer of PC representations is crucial in ensuring an accurate and effective reconstruction.
This is achieved through the use of branching blocks, which capture the core features and build the tree structure of the GCN. By doing so, the generator is capable of unsupervised preservation of the relative positional relationships between points and aggregation of original information from the sample of real-world . This approach helps to ensure that the reconstructed point clouds are as accurate and effective as possible.

\subsection{Stage-II GAN: upsampling with stack-structured GCN}
Once the stage-I GAN has generated a low-density PC, stage-II GAN can complete point cloud upsampling and reconstruct a high-density point cloud that can support more sophisticated brain-computer interface surgery. To achieve this, the basic GCN block is still used as the skeleton for building the stage-II generator, given its advantages for point cloud tasks. However, unlike the stage-I, the upsampling process in stage-II is more like a feature aggregation and re-generalization. Therefore, the geometrically designed tree-like network is not considered at this stage.
Compared with the advantages of tree-like GCN in image feature extraction and point cloud conversion expression, we believe that the second-stage GAN needs to have more powerful capabilities in subtle perception of local point set microstructure and hierarchical upsampling. At the same time, it should be able to control errors as much as possible while improving upsampling quality. Therefore, we considered a -like generator to gradually complete the second-stage reconstruction in a small but fine step. To conduct the point cloud upsampling and reconstruct a high-density PC with complex brain microstructure, the stage-II generator completes three steps: feature representation, upsampling and coordinate reconstruction.
%It is designed as a stack structure with three processes: feature extraction, feature upsampling, and coordinate reconstruction.
The low-density PC is updated in a sequential manner, using an information flow approach, until it is reconstructed back to a high-density point cloud. This process helps to ensure that the resulting point cloud is as accurate and realistic as possible, and can support more sophisticated brain-computer interface surgery.

The feature extraction component aims to extract features from the input $Y^l_{2048\times3}$ to a feature matrix $Y_{2048\times C}$ whose each column vector is extracted from the corresponding point of input. In this study, we utilize the feature extraction technique proposed in [38], which incorporates dense connections to integrate features across multiple layers.

The feature upsampling component is a super-resolution operator. It contains several basic GCN blocks and can expand node features $Y_{2048\times C}$ to shape $2048\times rC$. Then we apply a shuffle operation to rearrange the feature matrix. Finally, the output of this component expands to $Y_{2048r\times C}$.

The coordinate reconstruction component reconstructs points from latent space to coordinate space, resulting in the desired denser point cloud $Y^h_{2048r\times C}$. We construct a similar coordinate reconstruction approach as [17], in which 3D coordinates are regressed using two sets of FC-layers.

To complete the point cloud upsampling process and refine the details of the brain's point cloud representation, we use basic GCN blocks to design the components of a stack-structure GCN. This approach also allows us to correct any possible detail errors in the stage-I by using different loss. The illustration of generators for stage-I and stage-II can be seen in Figure 3. By using this approach, we are able to reconstruct a high-density PC that accurately represents the brain's microstructure.

\begin{algorithm}[htb]
\caption{ Framework of the proposed SG-GAN}
  \begin{algorithmic}[1]
    \Require
    $I_{H \times W}$: A single 2D MRI of brain;
    \Ensure
    $Y^{h}_{2048r \times 3}$: A high density point cloud;
    \State Initialize network parameters
    \For{iterator = 1, 2, 3, ..., T}
        \State Input $I_{H \times W}$ in free transforming module;
        \State Build parameter-free self-attention mechanism between layers through Eq.(5);
        \State Update features in each layer:
        \begin{equation}
        \bm{T^k} \gets softmax(A^k) \times \bm{T^k} \nonumber
        \end{equation}
        \State Generate latent feature vector $z \in \mathbb{R}^{96}$.
        \For{D\_iterator = 1, 2, 3, ..., t}
            \State Generate  $Y^{l}_{2048 \times 3}$ by using stage-I generator:
            \begin{equation}
            Y^{l}_{2048 \times 3} = \mathbb(G_1(z)) \nonumber
            \end{equation}
            \State Judge $Y^{l}_{2048 \times 3}$ and ground truth by using stage-I discriminator;
            \State Update parameters of stage-I discriminator;
        \EndFor
        \State Generate point cloud $Y^{l}_{2048 \times 3}$, judge $Y^{l}_{2048 \times C}$ and ground truth and Update parameters of stage-I generator;
        \For{D\_iterator = 1, 2, 3, ..., t}
            \State Feature extraction: extract  $Y^{l}_{2048 \times 3}$ to the feature matrix $Y_{2048 \times C}$;
            \State Feature upsampling: expand $Y_{2048 \times C}$ to the shape $2048 \times rC$, then shuffle:
            \begin{equation}
            Y_{2048r \times C} = Shuffle(Y_{2048 \times rC}) \nonumber
            \end{equation}
            \State Coordinate reconstruction: reconstruct $Y^{h}_{2048r \times 3}$ from latent space to coordinate space;
            \State Judge $Y^{h}_{2048r \times 3}$ and ground truth by using stage-II discriminator Update parameters of it;
        \EndFor
        \State Generate point cloud $Y^{h}_{2048r \times 3}$, judge and Update parameters of stage-II generator;
    \EndFor
    \State \Return $Y^{h}_{2048r \times 3}$ generated by well-trained network.
    \end{algorithmic}
\end{algorithm}

\subsection{Loss Functions in Training}
Herein, we introduce the compound loss function that has been specifically designed for training SG-GAN.

To better train GAN framework in an adversarial manner, we employ the similar strategies as WGAN-GP:
\begin{equation}
\mathcal{L}_{G} = - \mathbb{E}[D(G(f))],
\end{equation}

\begin{equation}
\begin{aligned}
\mathcal{L}_{D} =  \mathbb{E}[D(G(f))]& - \mathbb{E}_{Y\sim \mathcal{R}}[D(Y))]\\
& + \lambda_{gp}\mathbb{E}_{\hat{x}}[(||\nabla_{\hat{x}}D(\hat{x})||_2 - 1)^2],
\end{aligned}
\end{equation}
 here $f$ is the feature input, $\hat{x}$ are obtained by sampling along the line segments connecting the generated PC and the ground truth.
 $\mathcal{R}$ denotes the distribution of the samples from ground truth. And $\lambda_{gp}$ is a weight parameter used in the process. The stage-I generator and stage-I and II discriminators are trained by these loss functions.

We also want to guide the generator to better approximate the true value in the position of each point and make the reconstructed point cloud more reliable for each point, so we use the chamfer distance (CD):
\begin{equation}
\begin{aligned}
\mathcal{L}_{CD} = \sum\limits_{y' \in Y'} min_{y \in Y}& ||y' - y||^2_2 \\
& +  \sum\limits_{y \in Y} min_{y' \in Y'} ||y - y'||^2_2,
\end{aligned}
\end{equation}
here $y'$ and $y$ indicate the corresponding points in $Y'$ and $Y$, respectively.

Still, to promote the adherence of the generated points to the target surface, we incorporate a reconstruction loss based on the EMD index and it is given by:
\begin{equation}
\mathcal{L}_{EMD} = min_{\phi: Y \rightarrow Y'} \sum\limits_{x \in Y}||x - \phi(x)||_2
\end{equation}
here $\phi$ indicates a parameter of bijection.

Therefore, We give the loss function that guides the stage-II generator and encoder training:
\begin{equation}
\mathcal{L}_{U} = \lambda_1\mathcal{L}_{G} + \lambda_2\mathcal{L}_{KL} + \lambda_3\mathcal{L}_{CD} + \lambda_4\mathcal{L}_{EMD}.
\end{equation}
where $\lambda_1, \lambda_2, \lambda_3$ are weights.
The complete training process of the proposed SG-GAN is shown in Algorithm 1.

\section{Experiment}

\subsection{Experiment Preparation}

\subsubsection{Data Preparation}

This section describes the evaluations of the performance of the proposed SG-GAN through multiple experiments on an in-house dataset which comprises a total of 317 brain MRIs exhibiting Alzheimer's disease and 723 healthy brain MRIs. The dataset was preprocessed by removing all bone structures of brains and registering the remaining images into format 91 × 109 × 91 by the FSL software. 900 MRIs were randomly selected for the training set, and the remaining MRIs were used for the test set. The input of the model was chosen as some 2D slices of MRIs that were normalized to [0, 1]. Accurate voxel-level segmentation of all MRIs was performed, and the results were computationally transformed into point clouds.

\subsubsection{Implementation Details}

This section also describes the evaluation of the proposed model's performance by comparing it with several currently available advanced point cloud upsampling models, namely PU-net, PU-GAN, and 3PU, under the same conditions and scenarios. The experiments were conducted using Pytorch.
%on the CPU of Intel Core i9-7960X CPU @ 2.80GHz$\times$32 and GPUs of Nvidia GeForce RTX 2080 Ti. The values of $\lambda_1$, $\lambda_2$ and $\lambda_{gp}$ were set to 1, 0.1, and 10, respectively, and Adam optimizers were used with an initial learning rate of $1 \times 10^{-4}$. The training parameters were dynamically adjusted by increasing $\lambda_3$ from 0.1 to 1 and increasing $\lambda_4$ from 0.01 to 0.05. The experimental results are reported in the following.

\subsubsection{Evaluation Metrics}

We use some tools and metrics to evaluate the capacity and robustness of the SG-GAN. The effectiveness of the proposed method is assessed through the use of objective quality metrics, namely the Chamfer Distance (CD) and Earth Mover's Distance (EMD). Additionally, the PC-to-PC error is employed to quantify the error of each generated PC. The robustness of the proposed SG-GAN is evaluated using the Hausdorff distance (HD).

\begin{figure}[ht]
	\includegraphics[width=8.8cm]{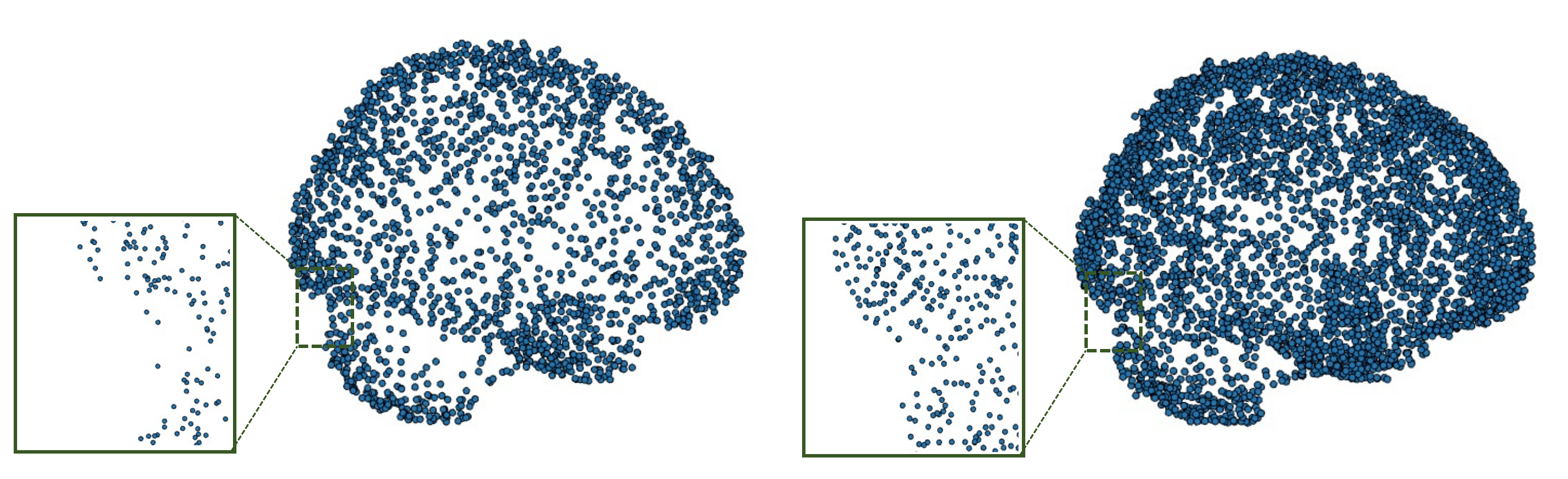}
	\caption{The comparison of microstructure in the same generating of the stage-I GAN and stage-II GAN.}
\end{figure}

\begin{figure}[ht]
	\centering
	\includegraphics[width=8.8cm]{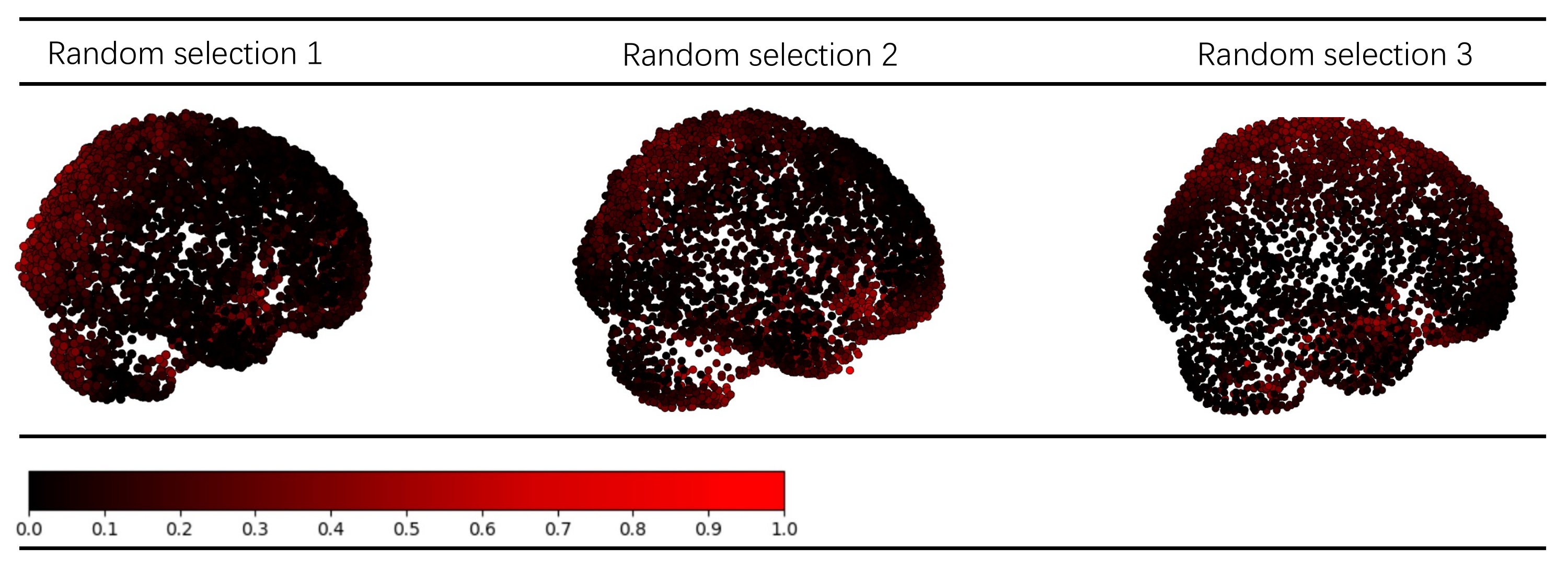}
	\caption {Reconstructed colored samples of 3D brain measured by PC-to-PC error. Heat map is used to show the exact value of the error.}
\end{figure}

\begin{figure}[ht]
	\centering
	\includegraphics[width=0.5\textwidth]{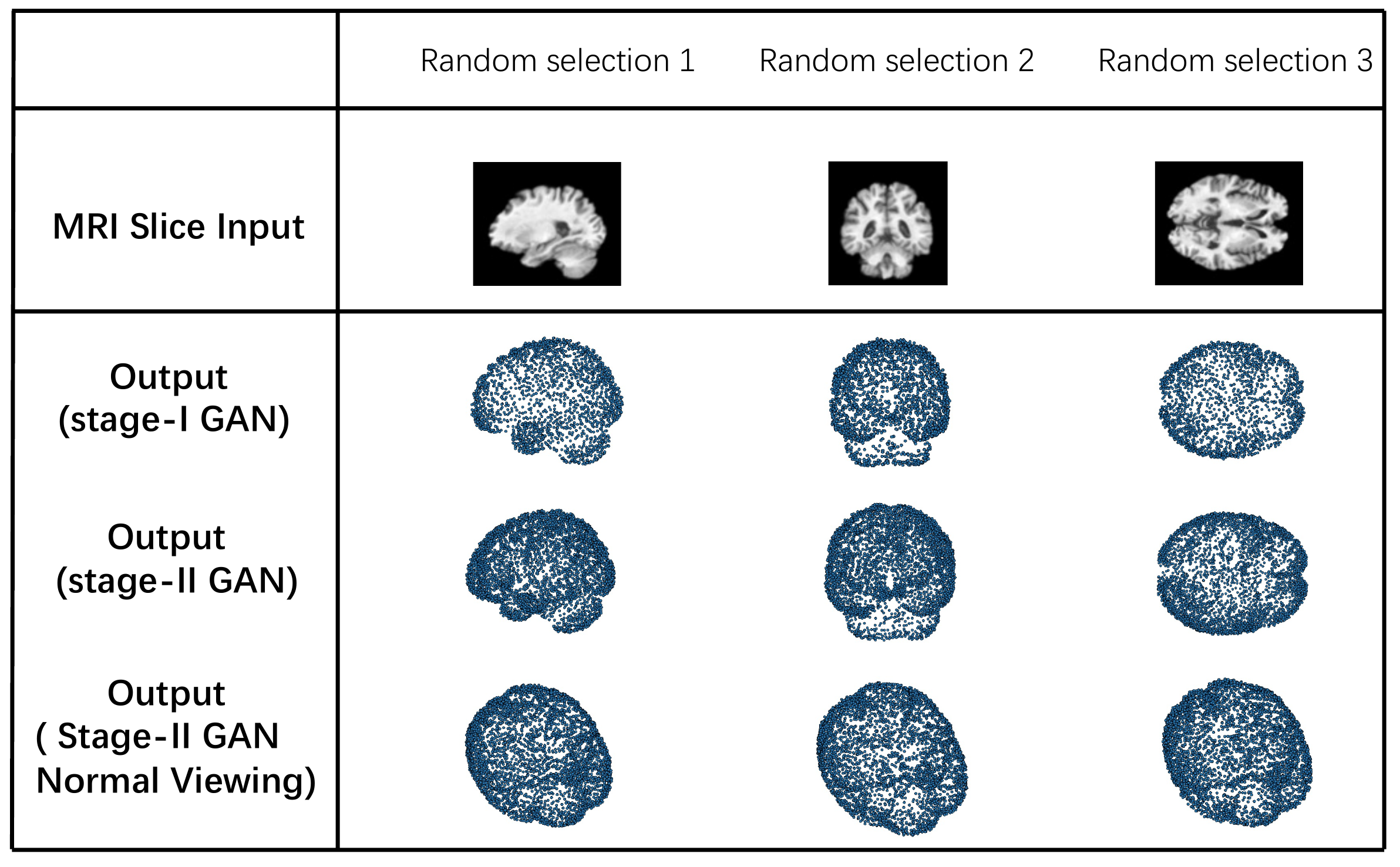}
	\caption{Qualitative evaluations of the proposed SG-GAN. Different angles of MRI are chosen as inputs for evaluating the SG-GAN. The reconstructed PC is reported by showing the results of stage-I GAN and stage-II GAN in the 3D space and projecting them onto the normal viewing.}
\end{figure}

\begin{figure}[ht]
	\centering
	\includegraphics[width=8.8cm]{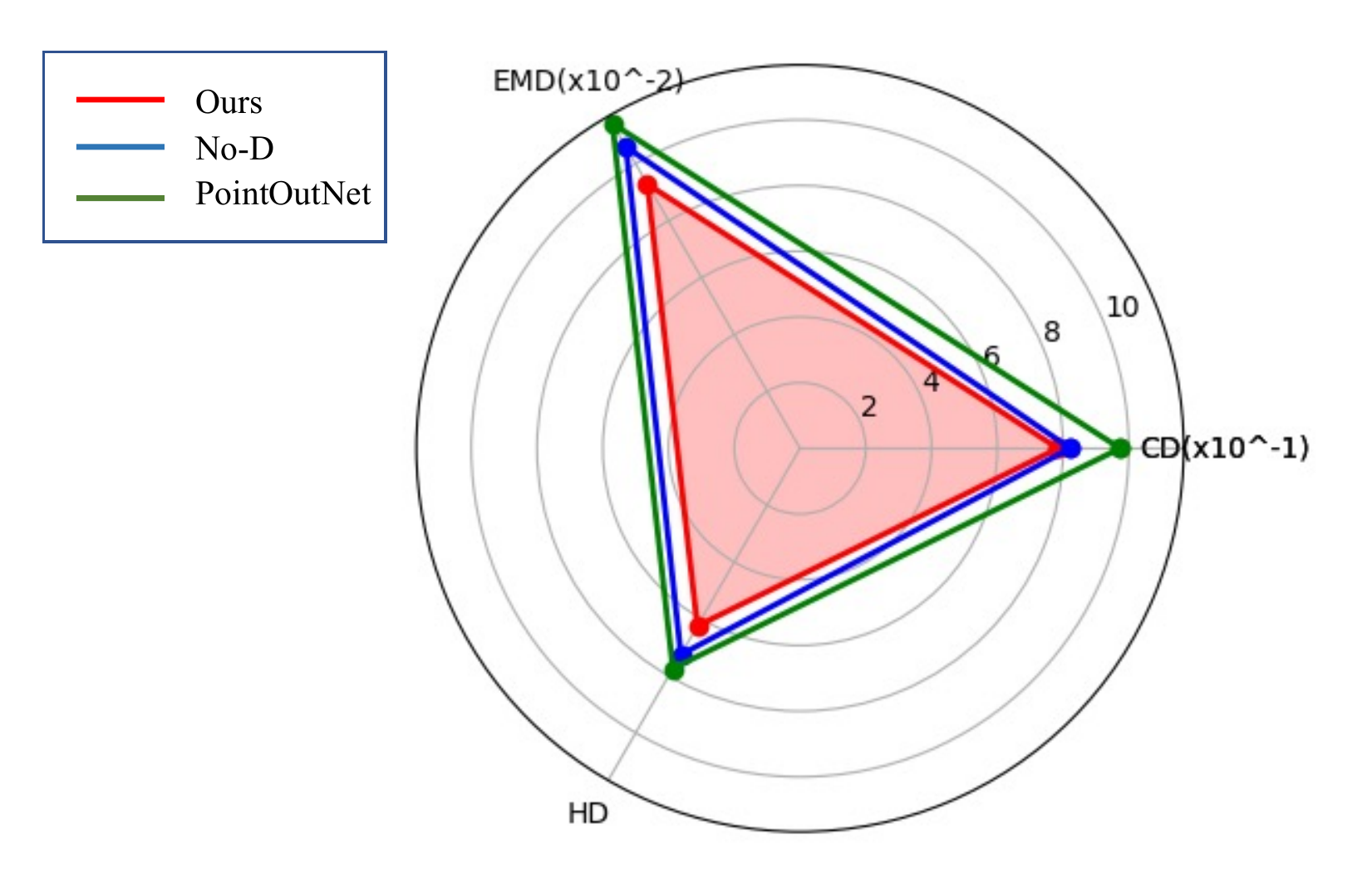}
	\caption{The effect of different stage-I generating systems shown by radar chart.}
\end{figure}

We performed a comprehensive analysis of the SG-GAN model on the constructed dataset. Taking into account the complexity of the task accomplished by the proposed approach, and the fact that most point cloud upsampling methods do not have the ability to extract image features, we build the same encoder as SG-GAN for all the methods involved in the comparison.

\subsection{Completion Performance Analysis}
Some experimental results of the proposed SG-GAN on the test set and qualitative evaluations are given in Fig. 4.
In particular, we present together the intermediate state generation results of the stage-I generator to show the perceptive ability and detail-filling ability of the proposed GAN for the overall shape of the brain. Figure 5 shows the comparison of microstructure in the same generating of the stage-I GAN and stage-II GAN. An analysis of the point-by-point error of generated point clouds is shown in Fig. 6. To reflect the difference point by point between the reconstructed point clouds and their corresponding ground truth, a difference heatmap is employed. kinds of PC angles are selected to show the advantage using the heatmap, and to show the accurate reconstruction. These experimental results demonstrate that the proposed approach exhibits a high level of sensitivity to detail, as the majority of vertices in the reconstructed point clouds exhibit very small errors. It is worth noting that the values on the axis have been scaled by a factor of $10^{-4}$.

\begin{figure*}[ht]
	\centering
	\includegraphics[width=0.9\textwidth]{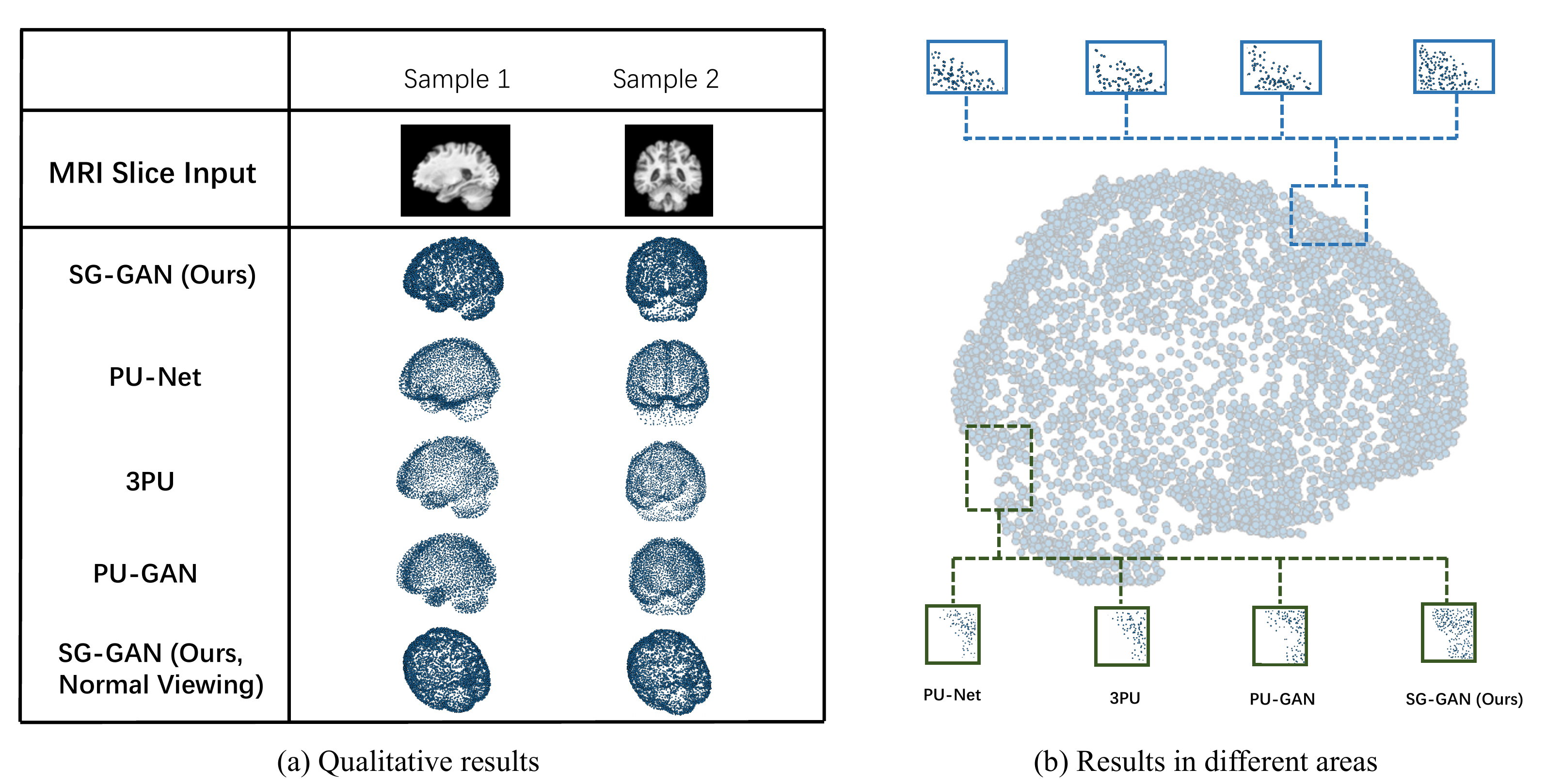}
	\caption{Comparing the proposed SG-GAN with several existing methods. (a) Qualitative analysis using MR slices of different angles. (b) The reconstructed PC evaluation in terms of different ROIs.}
\end{figure*}

\subsection{Ablation Study}

To evaluate the proposed SG-GAN, we examined how crucial modules and hyper-parameters impact SG-GAN. For ease of convenience, all research is usually carried out on the brain category. We assess the overall quality of point cloud generation by using CD, EMD, and HD metrics. To account for variations in the number of generated points in specific different regions, we employ the error of PC-to-PC to evaluate each reconstructed point.

\begin{figure*}[ht]
	\centering
	\includegraphics[width=0.9\textwidth]{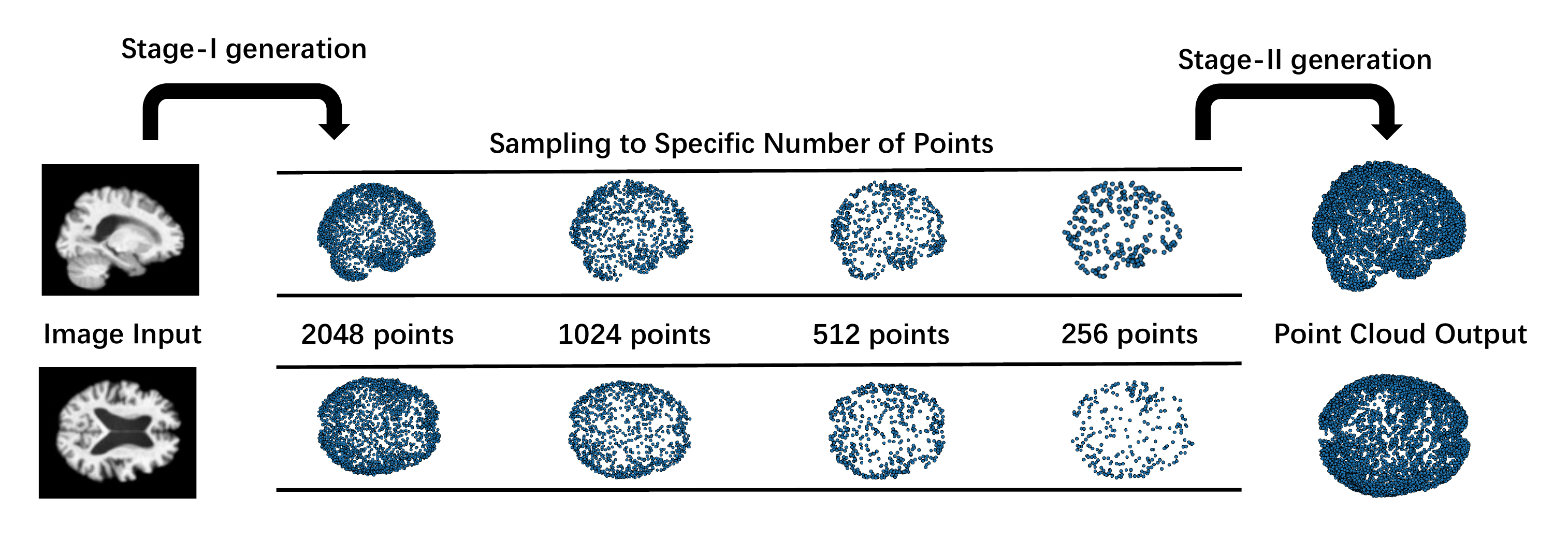}
	\caption{Comparison of different point numbers of stage-I GAN. Points are randomly sampled via a predetermined number from 2048 points and then input into the stage-II GAN.}
\end{figure*}

\subsubsection{The performance of Encoder}
To demonstrate the efficiency and precision of the proposed FTM-based encoder, we conducted several modifications to the encoder and evaluated the error. (i) ``ResNet" often indicates a variation which eliminates the attention calculation, thereby removing the self-attention scheme of the encoder network. (ii) ``NormalNet" indicates a variation that employs the standard attention network to replace the parameter-free self-attention calculation structure. The comparison of overall point error results is presented in Table I, confirming that our designed module is the most effective in aiding the model to achieve precise reconstruction targets.
\begin{table}[htbp]
	\centering
	\caption{The effect of different encoder structure measured by CD$(\times10^{-1})$}
	\begin{tabular}{p{2cm}|ccc}
		\toprule
		Method&Ours&ResNet&NormalNet \\
		\midrule
		CD$(\times10^{-1})$&\color{red}{7.852}&\color{blue}{8.815}&8.114 \\
		EMD$(\times10^{-2})$&\color{red}{9.285}&\color{blue}{10.271}&9.963 \\
		HD&\color{red}{6.247}&\color{blue}{6.958}&6.709 \\
		\bottomrule
	\end{tabular}
\end{table}

\begin{table}[htbp]
	\centering
	\caption{The effect of different stage-I generating system}
	
	\begin{tabular}{p{2cm}|ccc}
		\toprule
		Method&Ours&No-D&PointOutNet\\
		\midrule
		CD$(\times10^{-1})$&\color{red}{7.852}&8.225&\color{blue}{9.741} \\
		EMD$(\times10^{-2})$&\color{red}{9.285}&10.592&\color{blue}{11.413}\\
		HD&\color{red}{6.247}&7.251&\color{blue}{7.740}\\
		\bottomrule
	\end{tabular}
\end{table}

\subsubsection{Effect of Stage-I Generating System}
The generative capacity of neural networks has a direct impact on the fitting performance of models and ultimately determines the quality of the generated point clouds. To assess this capability, we replaced the generative system (stage-I and stage-II GAN) with several different models and tested their performance. In particular, since some of the models we compared are only point cloud-generating models, we also added FTM-based encoders that are the first to the proposed models for other comparison models to enable them to obtain the same ability to extract image features. In this section, we conducted two modifications to our stage-I GAN and evaluated the performance. (i) ``No-D" indicates a variation which eliminates the corresponding discriminator of the stage-I GAN, thereby removing the adversarial learning effect of the GAN structure. Besides, we solely employ Chamfer distance to train the remaining network. (ii) ``PointOutNet" is a variation that employs the PointOutNet in [48], to substitute the proposed stage-I GAN framework. The experimental results, presented in Table II and Fig. 7, confirm that the proposed structure is the most effective in aiding the model to achieve precise construction targets.

\begin{table}[htbp]
	\centering
	\caption{The effect of stage-II different generating system}
	\setlength{\tabcolsep}{2mm}{
		\begin{tabular}{p{2cm}|ccccc}
			\toprule
			Method&Ours&one-stage&PU-Net&3PU&PU-GAN\\
			\midrule
			CD$(\times10^{-1})$&\color{red}{7.852}&17.458&\color{blue}{18.247}&17.074&16.551 \\
			EMD$(\times10^{-2})$&\color{red}{9.285}&18.741&\color{blue}{20.858}&19.060&18.527\\
			HD&\color{red}{6.247}&15.034&17.277&\color{blue}{17.371}&16.586 \\
			\bottomrule
	\end{tabular}}
\end{table}

\begin{figure}[htbp]
	\centering
	\includegraphics[width=0.5\textwidth]{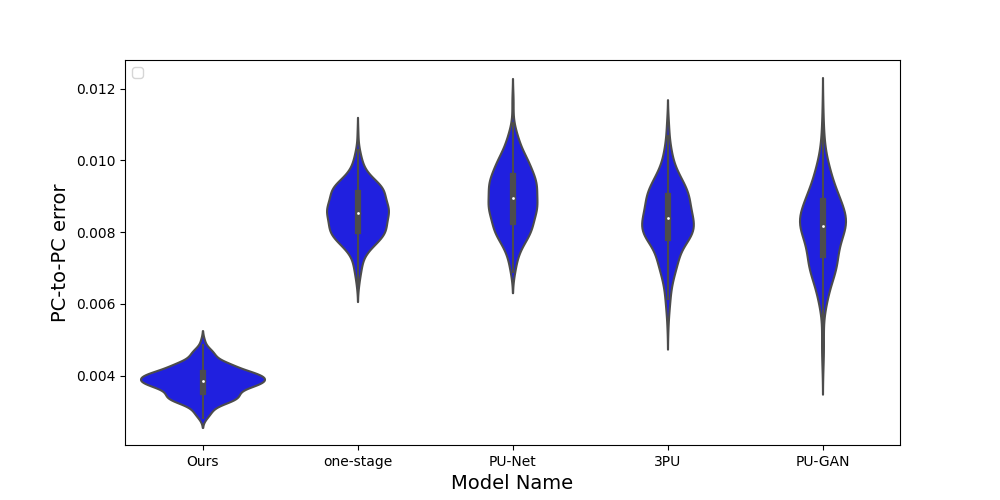}
	\caption{The violin plot of the PC-to-PC error results of the Stage-II Generating System ablation study.}
\end{figure}

\subsubsection{Effect of Stage-II Generating System}

After the stage-I GAN, we set up the upsampling stage-II GAN that upsamples low-density point clouds into high-density point clouds. In this section. We want to demonstrate the following two points: a) A two-stage generation network like ours is superior to a single one-stage generation system for high-density point clouds. b) The model structure of our stage-II GAN is superior to other current SOTA work. Therefore, we set up the following experiment, we also added an FTM-based encoder to these experiments for other comparison models : (1) ``one-stage" is the variation that only uses stage-I tree-structured graph convlution GAN to generat high-density PC directly. (2) ``PU-Net" indicates the variation which conducts a PU-Net based stage-II generator. (3) ``3PU" is to replace the stage-II generator with 3PU. (4) ``PU-GAN" is to replace the stage-II generator with PU-GAN. In Fig. 8, we present the reconstruction process for two randomly selected samples from the test set, with the corresponding detailed results reported in the right part of the figure and Table III. The comparisons of total errors and mean provide evidence of the superior reconstruction ability of our proposed model.

At the same time, compared with the evaluation indicators which measure the overall error of the point cloud such as CD and EMD, the PC-to-PC error better evaluates the similarity between the single point in the point cloud and the real sample. We give a schematic representation of the a-errors for these sets of experiments in Fig. 9. The results show that our generated results also have the best performance on the single point.

\subsubsection{Effect of Point Input Size}

In this study, the point cloud output size of our stage-I GAN (also, the input size of the stage-II GAN) is 2048, and the model exhibits good performance at this scale. Additionally, we want to investigate whether the model's performance remains consistent when input conditions deteriorate. To test the robustness of the stage-II generator, we randomly selected n points from the outputs of the stage-I GAN and transferred them to the stage-II generator. This was carried out to ensure that the visual and sensing capabilities during surgery do not affect the generation efficiency. We conducted tests with multiple values of n, as explained in Fig. 10. The experimental results are presented in Table IV.

\begin{table}[htbp]
	\centering
	\caption{The effect of different point numbers}
	\setlength{\tabcolsep}{5mm}{
	\begin{tabular}{p{2cm}|ccc}
		\toprule
		point numbers&2048&1024&512 \\
		\midrule
		CD$(\times10^{-1})$&\color{red}{7.852}&7.923&\color{blue}{8.159} \\
		EMD$(\times10^{-2})$&\color{red}{9.285}&9.745&\color{blue}{9.768} \\
		HD&\color{red}{6.247}&6.441&\color{blue}{6.672} \\
		\bottomrule
	\end{tabular}}
\end{table}

\begin{figure*}[ht]
	\centering
	\includegraphics[width=18cm]{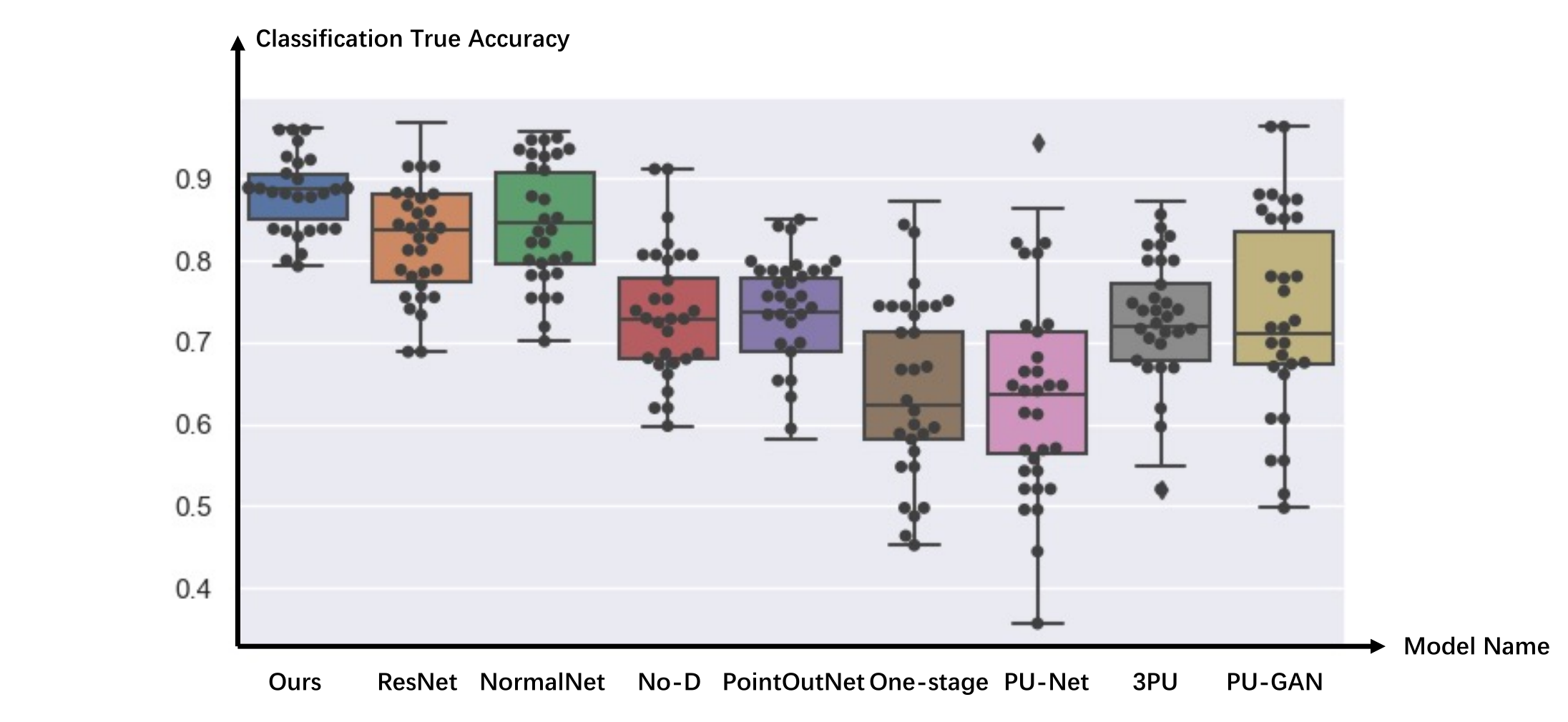}
	\caption{ Comparison on classification performance using the reconstructed point clouds. The classification accuracy is compared between the proposed SG-GAN and the other methods, including ResNet, NormalNet and PU-Net et al. The experimental results shows that the point clouds generated by the proposed SG-GAN represent more effective features of ground truth.}
\end{figure*}

\subsubsection{Effect of Input Number}

In this study, we utilized a single slice of MR to predict the 3D high-density point clouds of the target brains, and we demonstrated the high accuracy of the reconstructed point cloud. However, it is still meaningful to study the relationship between the number of input images and the quality of generation. Our hypothesis was that utilizing multiple MRI slices as input to the model, as opposed to a single image, would enhance the feature aggregation via the encoding structure, ultimately leading to strengthening the quality of PC reconstruction via decoding process. To test the hypothesis, we designed an experiment in which multiple slices of MRI are employed as input to reconstruct 3D brain PC. The experimental results, presented in Table V, indicate that although using multiple slices as input does improve reconstruction accuracy, the degree of improvement is not significant enough to justify the necessity to acquire it by an increase in computational complexity associated with processing multiple images. Therefore, we conclude that one single image is an appropriate input size for the specific scenario studied in this work.

\begin{table}[htbp]
	\centering
	\caption{The effect of different input numbers}

	\begin{tabular}{p{1.7cm}|cccc}
		\toprule
		input numbers&single slice&3 slices&5 slices&7 slices \\
		\midrule
		CD$(\times10^{-1})$&7.852&7.712&7.704&\textbf{7.667} \\
		EMD$(\times10^{-2})$&9.285&9.109&9.080&\textbf{9.042} \\
		HD&6.247&6.175&6.122&\textbf{6.083} \\
		\bottomrule
	\end{tabular}
\end{table}

\subsection{Evaluation using Classification Tasks}

Evaluation metrics for reconstructed point clouds do not provide an intuitive evaluation of their visual acceptability, although they can demonstrate the similarity of the output and ground truth. We would like to further demonstrate that the reconstructed results of the proposed model are also more cognitively plausible and closer to the real visual effect. To study this task, We assessed the classification similarity of point clouds generated by our SG-GAN and the models constructed in the ablation study utilizing the PointNet++ classifier network in the context of a binary classification task. To ensure a fair comparison, we trained our classifier by using the same training strategy employed in the discriminator training.
%First, we performed pre-training of the PointNet++ model using two data-sets of PC. the PC data indicating ground truth, labeled as 'true', and the point clouds generated by each method trained up to 1000 epochs, labeled as 'false'. Then, we input the generated point clouds from different methods in the test to the classifier and compared their ability to fool the classifier through a classification experiment.
The experimental results are presented in Fig. 11. Bases on Fig.11, it can be concluded that the proposed SG-GAN outperformed the other reconstruction methods in terms of its ability to fool the classifier. These experimental results indicate that the point clouds generated by SG-GAN exhibit superior quality and realism, and are able to restore a greater amount of microstructural information pertaining to the real brain features.

\section{Discussion and Conclusion}
In this paper, A novel image-to-PC reconstruction network named SG-GAN is proposed to tackle the problem of low definition and density of existing 3D reconstruction methods in BCI minimally invasive surgeries. The proposed SG-GAN enables 3D brain reconstruction technology from which surgical navigation can benefit.
Given the need for real-time information feedback in surgical navigation and the associated computational-time-cost, we have opted to utilize point clouds as the representation for our proposed model, with one single image serving as the input.

An unsophisticated alternative to the proposed model is training a basic model and reconstruct the entire high-density point clouds from the input in a single step. Nevertheless, such a alternative may be impractical due to the following two aspects.  (1) The prior knowledge and known information from original image can not be fully exploited; (2) some microstructures of the generated high-density point cloud are not adequately adjusted in geometric angle, which can result in greater detail errors.

By integrating several complementary modules, the proposed SG-GAN enables the completion of scheduled reconstruction of complex brain shape.
 to jointly complete scheduled complex reconstructing and upsampling tasks. A parameter-free attention mechanism based on numerical computation is designed to constitute the encoder of the proposed GAN. This encoder adjusts the weights of the extracted features using spatial domain self-attention while ensuring the temporal sensitivity of the surgical scene, thus making the output feature vector more shape plausible. And then, two different stages of GANs are designed to form the generative network. Stage-I GAN focuses on outlining the specific shape of the target brain and generates low-density point clouds using the features. Stage-II GAN focuses on depicting the specific details of the target brain and fixes some of generation errors from the stage-I to reconstruct the final high-density PC.

Currently, intraoperative MRI technology has made significant improvement for surgical navigation. In this study, we utilized a single slice of brain MR image as input to reconstruct the corresponding 3D brain shape and the obtain results are promising. Our goal is to provide doctors with immediate access to 3D brain shape with high-density during brain surgical navigation. For the future work, we will collaborate with clinicians and collect data produced in brain surgical navigation from real-world so that we can remove the input constraints of the proposed model as much as possible. Finally, our model can be adapted to the input of the real-world dataset and achieve the same competitive 3D shape structure.

%Given that some of the existing state-of-the-art methods do not align with our experimental objectives, we incorporated additional modules into some other models in our comparison experiments to suit the current scenario. The experimental design confirms the validity and reliability of our models. In both qualitative and quantitative experiments, our model outperforms other state-of-the-art combinations. The classification experiments demonstrate that the generative structure of our model is more "realistic" in the solution space. Our approach has a very short inference time and can provide real-time feedback on local image properties, which can guide physicians to the diagnostic value of the surgical location effectively.

\section*{Acknowledgment}
This work was supported by National Natural Science Foundations of China under Grant 62172403 and 12326614, and the Distinguished Young Scholars Fund of Guangdong under Grant 2021B1515020019. M. Ng was supported by the National Key Research and Development Program of China under Grant 2024YFE0202900, HKRGC GRF 17300021, C7004-21GF and Joint NSFC-RGC N-HKU76921.

\bibliographystyle{IEEEtran}
\bibliography{ref}

\end{document}